\documentclass{emulateapj}

\newcommand{\kms}{\rm km \ s^{-1}}
\newcommand{\kmsmpc}{\rm km \ s^{-1} \ Mpc^{-1}}

\slugcomment{}

\shorttitle{IR Light Curve of SN 2011fe}
\shortauthors{Matheson et al.}

\begin{document}


\title{The Infrared Light Curve of SN 2011fe in M101 and the Distance
  to M101}

\author{T. Matheson\altaffilmark{1}, 
R.~R. Joyce\altaffilmark{1}, 
L.~E. Allen\altaffilmark{1},
A. Saha\altaffilmark{1}, 
D.~R. Silva\altaffilmark{1},
W.~M. Wood-Vasey\altaffilmark{2},  
J.~J. Adams\altaffilmark{3}, 
R.~E. Anderson\altaffilmark{4},
T.~L. Beck\altaffilmark{4},
M.~C. Bentz\altaffilmark{5},
M.~A. Bershady\altaffilmark{6},
W.~S. Binkert\altaffilmark{1},
K. Butler\altaffilmark{1},
M.~A. Camarata\altaffilmark{7},
A. Eigenbrot\altaffilmark{6},
M. Everett\altaffilmark{1},
J.~S. Gallagher\altaffilmark{6},
P.~M. Garnavich\altaffilmark{8}
E. Glikman\altaffilmark{9},
D. Harbeck\altaffilmark{10},
J.~R. Hargis\altaffilmark{11},
H. Herbst\altaffilmark{6},
E.~P. Horch\altaffilmark{7},
S.~B. Howell\altaffilmark{12},
S. Jha\altaffilmark{13},
J.~F. Kaczmarek\altaffilmark{6},
P. Knezek\altaffilmark{10,1},
E. Manne-Nicholas\altaffilmark{5},
R.~D. Mathieu\altaffilmark{6},
M. Meixner\altaffilmark{4},
K. Milliman\altaffilmark{6},
J. Power\altaffilmark{1},
J. Rajagopal\altaffilmark{1},  
K. Reetz\altaffilmark{1},
K.~L. Rhode\altaffilmark{11},
A. Schechtman-Rook\altaffilmark{6},
M.~E. Schwamb\altaffilmark{9},
H. Schweiker\altaffilmark{10},
B. Simmons\altaffilmark{9},
J.~D. Simon\altaffilmark{3},
D. Summers\altaffilmark{1},
M.~D. Young\altaffilmark{11},
A. Weyant\altaffilmark{2},
E.~M. Wilcots\altaffilmark{6},
G. Will\altaffilmark{1},
D. Williams\altaffilmark{1}}

\altaffiltext{1}{National Optical Astronomy Observatory, 950 N. Cherry
  Avenue, Tucson, AZ 85719} 
\altaffiltext{2}{Pittsburgh Particle Physics, Astrophysics, and Cosmology Center (PITT-PACC),
University of Pittsburgh, 3941 O'Hara Street, Pittsburgh, PA 15260, USA}
\altaffiltext{3}{Observatories of the Carnegie Institution of Washington, 813 Santa Barbara Street, Pasadena, CA 91101, USA}
\altaffiltext{4}{Space Telescope Science Institute, 3700 San Martin
  Drive, Baltimore, MD 21218 USA} 
\altaffiltext{5}{Department of Physics and Astronomy, Georgia State University, Astronomy Offices, One Park Place South SE, Suite 700, Atlanta, GA 30303, USA}
\altaffiltext{6}{Department of Astronomy, University of Wisconsin-Madison, 475 N. Charter Street, Madison, WI 53706, USA}
\altaffiltext{7}{Department of Physics, Southern Connecticut State University, 501 Crescent Street, New Haven, CT 06515, USA}
\altaffiltext{8}{Department of Physics, University of Notre Dame, Notre Dame, IN 46556, USA}
\altaffiltext{9}{Department of Astronomy, Yale University, New Haven, CT 06511, USA}
\altaffiltext{10}{WIYN Observatory, 950 N. Cherry Avenue, Tucson, AZ
  85719 USA}
\altaffiltext{11}{Department of Astronomy, Indiana University, Bloomington, IN 47405, USA}
\altaffiltext{12}{NASA Ames Research Center, Moffett Field, CA 94035}
\altaffiltext{13}{Deptartment of Physics and Astronomy, Rutgers, The State University of New Jersey, Piscataway, NJ 08854-8019, USA}

\email{matheson@noao.edu}

\begin{abstract}
We present near infra-red light curves of supernova (SN)~2011fe in
M101, including 34 epochs in $H$ band starting fourteen days before
maximum brightness in the $B$-band.  The light curve data were
obtained with the WIYN High-Resolution Infrared Camera (WHIRC).  When
the data are calibrated using templates of other Type Ia SNe, we
derive an apparent $H$-band magnitude at the epoch of $B$-band maximum
of $10.85 \pm 0.04$.  This implies a distance modulus for M101 that
ranges from 28.86 to 29.17 mag, depending on which absolute
calibration for Type Ia SNe is used.
\end{abstract}


\section{Introduction}

Supernova 2011fe was discovered soon after explosion in the nearby
galaxy M101 (NGC~5457) by the Palomar Transient Factory
\citep{rau09,law09} on 2011 August 24.167 UT (all calendar dates
herein are UT) and rapidly classified as a supernova of Type Ia (SN
Ia) \citep[][identified initially as PTF11kly]{nugent11a, nugent11b}.
This was the brightest SN Ia since SN~1972E in NGC~5253 \citep[see,
  e.g.,][]{kirshner73}, although SN~1986G remains the nearest SN~Ia
\citep[e.g.,][]{phillips87}.  The combination of proximity, early
discovery, and modern observing resources makes this SN a rare gift
that can be studied in unprecedented detail.

Over the last two decades, the reliability of SNe~Ia as calibratable
standard candles has been securely established. Specifically, the
regression of peak brightness in optical magnitudes (corrected for
reddening and \emph{a priori} second parameter effects, usually
light-curve shape) against recession velocity $v_r$ (the Hubble
diagram) in the range $1200\ \kms \leq v_{r} \leq 30,000\ \kms $ has
been shown to be almost linear and to have a scatter of only 0.13 mag
rms \citep[e.g.,][]{hamuy96,riess96, parodi00, guy05,
  jha07,conley08,hicken09,kessler09, folatelli10,burns11, mandel09,
  mandel11}. SNe~Ia are thus largely free of any Malmquist bias, and
by virtue of their brightness, are cosmological probes at distances
where peculiar motions of galaxies are insignificant compared to the
expansion velocity of the cosmic manifold. Their consequent use as
probes of cosmic acceleration is now well known
\citep[e.g.,][]{riess98,perlmutter99,astier06,woodvasey07,sullivan11}.

Photometry of SNe~Ia in the near infra-red (NIR) has shown even
greater promise for use as a standard candle (for a recent review of
the NIR properties of SNe~Ia, see \citet{phillips11}).  The effects of
extinction are greatly reduced and they appear to have relatively
constant peak magnitudes in $J, H,$ and $K_s$ \citep{meikle00,
  krisciunas04a, krisciunas04b, krisciunas07}.  The scatter in the NIR
Hubble diagram is $\sim 0.15$ mag \emph{without} the light-curve shape
corrections necessary for optical bands \citep{krisciunas04a,
  woodvasey08, folatelli10}.  Motivated by this, we began a Director's
discretionary time program to observe SN~2011fe with the WIYN
High-Resolution Infrared Camera (WHIRC) NIR camera at the WIYN 3.5-m
telescope\footnote{The WIYN Observatory is a joint facility of the
  University of Wisconsin, Indiana University, Yale University, and
  the National Optical Astronomy Observatory.} on Kitt Peak, taking
advantage of the instrumentation deployment that allows for the use of
the NIR camera even when other instruments are scheduled for a given
night.

\section{Observations and Data Reduction}


We imaged the SN on 34 nights between 2011 Aug 27 and 2011 Oct 26 with
WHIRC.  The WHIRC camera \citep{meixner10} contains a 2048$^2$ HgCdTe
array with a field of view of 3\farcm3 x 3\farcm3 and a pixel scale of
$\sim$0.1\arcsec\ per pixel.  We obtained observations in the $H$ band
during each visit and observations in the $J$ and $K_s$ bands for most
of the visits (see Table \ref{irphot} for details).  On photometric
nights, a standard star at similar air mass
\citep[P133C,][]{persson98} was observed in the same filters.  A
typical observing sequence consisted of 20 to 30 second exposures in a
5-point cross-shaped dither pattern with
$\sim$20\arcsec\ dithers. Most nights, this sequence was executed
twice in the $H$ band, with a 5\arcsec\ random offset of the telescope
between maps.

Data were reduced in IRAF\footnote{IRAF is distributed by the National
  Optical Astronomy Observatory, which is operated by the Association
  of Universities for Research in Astronomy, Inc., under cooperative
  agreement with the National Science Foundation.} as prescribed in
the WHIRC Reduction Manual \citep{joyce09}.The raw images were
corrected for non-linearity and sky subtracted using a median-filtered
sky frame obtained from each 5-point map.  The images were
flat-fielded with dome flats corrected for the pupil ghost (an
inherent feature of WHIRC images resulting from internal reflections
from the optical elements) using the IRAF routine mscred.rmpupil.
Aperture photometry was performed on each sky-subtracted, flatfielded
image using typical apertures of 3\arcsec\ diameter and a surrounding
sky annulus of 0.5\arcsec\ width.  On nights where the seeing FWHM was
greater than 1.0\arcsec, a 4 or 5\arcsec\ aperture diameter was used.
We did not have a template image to subtract the host galaxy light as
is commonly done for SNe.  The host background is relatively smooth in
the region chosen for the sky.  In addition, inspection of Two Micron
All Sky Survey \citep[2MASS;][]{skrutskie06} images of the region
reveal no point source at the location of SN~2011fe to 3$\sigma$
limits of 17.5, 17.4, and 16.7 in $J, H,$ and $K_s$, respectively,
well below the brightness of the SN itself.

\begin{figure}
\epsscale{.80}
\plotone{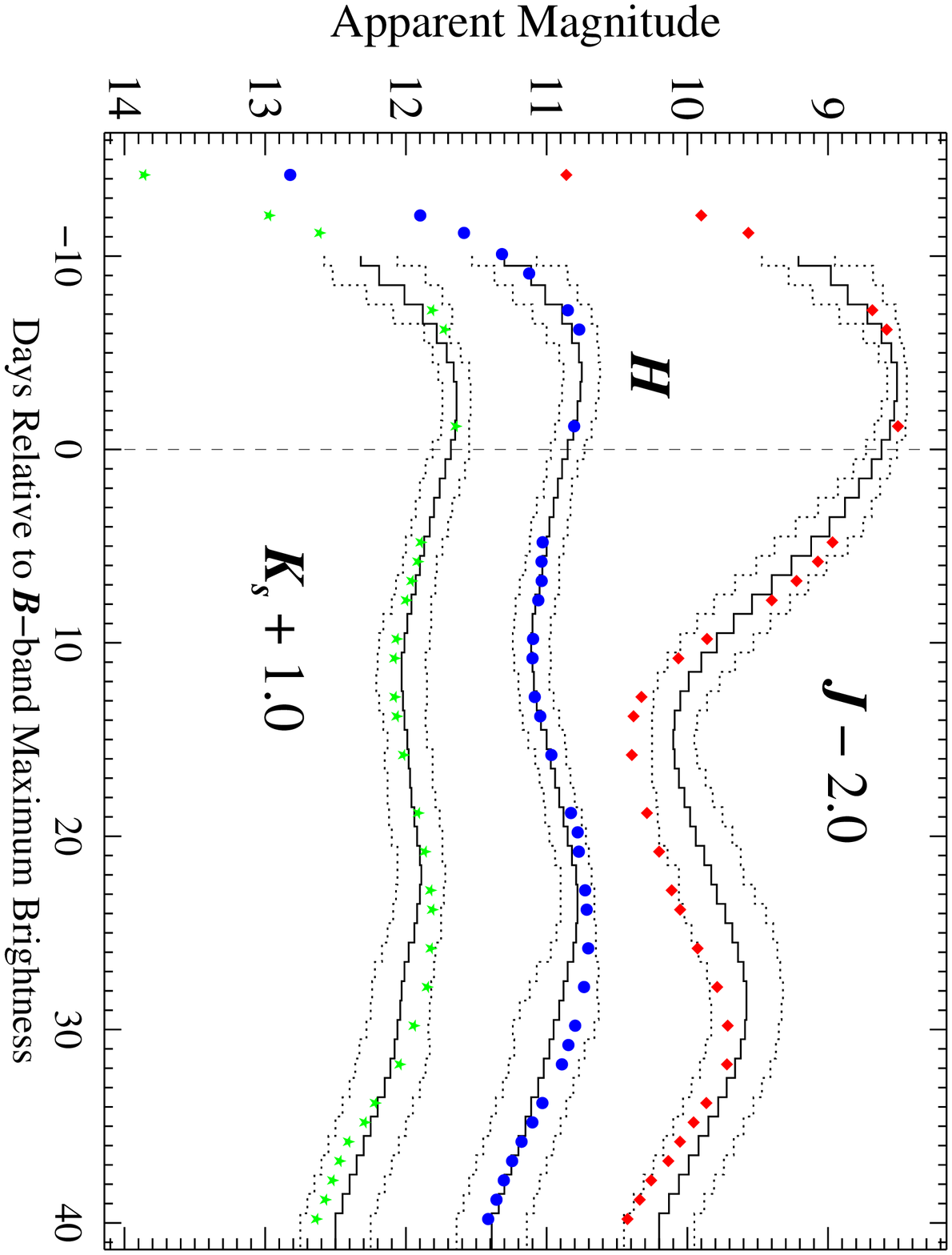}
\caption{Light curves of SN~2011fe in $J$ (red diamonds), $H$ (blue
  circles), and $K_s$ (green stars).  Error bars are smaller than the
  plotted symbols.  The $J$-band points are offset by -2.0 mag while
  the $K_s$-band points are offset by +1.0 mag.  The templates of
  \citet{woodvasey08} along with their 1-$\sigma$ envelopes are
  plotted for each passband.  Note that our $J$ filter is
  significantly different than the 2MASS $J$ filter used for the
  \citet{woodvasey08} template.  The vertical dashed line marks the
  epoch of $B$-band maximum.\label{lcfig}}
\end{figure}

All photometry of SN~2011fe was calibrated relative to a local
calibrator, the nearby 2MASS source 14031367+5415431. This star lies
approximately 80\arcsec\ SW of SN~2011fe, within the same WHIRC field,
so the relative photometry should be independent of atmospheric
transparency or extinction.  Because of the relatively large (0.024
mag) published uncertainties in the 2MASS calibration, the nearby
photometric standard P133C \citep{persson98} was observed on ten
photometric nights to calibrate the local 2MASS standard using
canonical atmospheric extinction coefficients of 0.08, 0.04, and 0.07
mag/airmass for the $J$, $H$, and $K_s$ filters, respectively.  The
corrections to the 2MASS magnitudes of the local calibrator to the
Persson standard were small ($\sim 0.01$ mag), with our final values
being $12.008 \pm 0.009$ mag, $11.471 \pm 0.008$ mag, and $11.405 \pm
0.011$ mag in $J$, $H$, and $K_s$, respectively.  The flux-calibrated
magnitudes of SN~2011fe are presented in Table \ref{irphot} and
plotted in Figure \ref{lcfig}.  The quoted uncertainties are the rms
of the mean of the individual images with the error in the calibration
of the standard (which is the dominant source of error) included.

In addition, the brightness of the local standard 2MASS
14031367+5415431 was measured relative to another 2MASS star in the
field (14025941+5416266) on all nights to ensure against intrinsic
variability of the local standard.  Over the course of the
observations, the rms uncertainty in the differential photometry of
these two stars was 0.015 mag, which can be considered an upper limit
on any intrinsic variability.  If nights judged not to be photometric
are excluded, this rms uncertainty decreases to 0.010 mag.

\section{Analysis}

\begin{figure}
\epsscale{.80}
\plotone{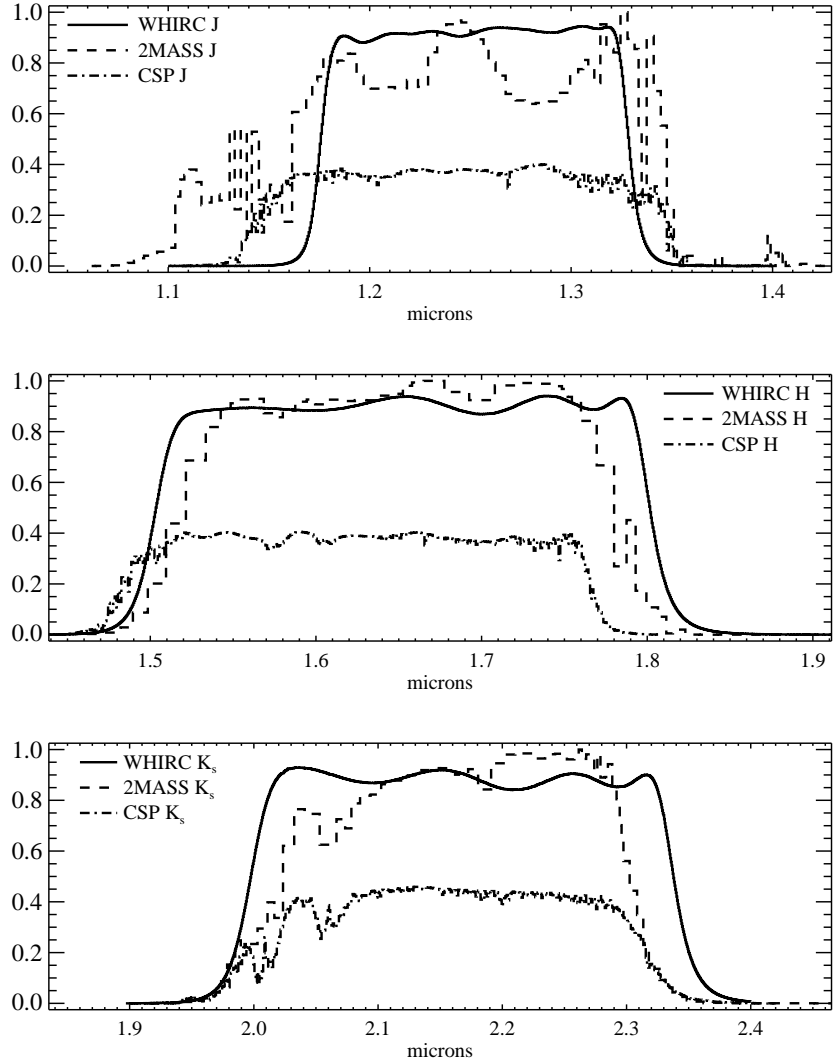}
\caption{Filter bandpasses for the $J$, $H$, and $K_s$ bands for the
  2MASS filters used for the PAIRITEL data
  \citep{woodvasey08}, the filters used for CSP data
  \citep{contreras10}, and the WHIRC filters.  For 2MASS and CSP, the
  bandpasses reflect the total system throughput.  Only filter
  transmission is available for the WHIRC filters.\label{filtfig}}
\end{figure}

In order to calibrate SN~2011fe against other SNe~Ia, we compared our
photometry with that of \citet{woodvasey08}.  \citet{kattner12} have
shown that there is a weak relationship between absolute luminosity
and decline rate in the $J$ and $H$ bands.  Optical photometry of
SN~2011fe indicates that it is a ``normal'' SN Ia with a $\Delta
m_{15}(B)$ value of $\sim$1.2 \citep{richmond12}, so we will not make
any corrections for decline rate (as the correction suggest by
\citet{kattner12} is minimal at this decline rate).
\citet{woodvasey08} provide templates of light curves of SNe~Ia in $J,
H, $ and $K_s$.  We fit our data to the templates using a $\chi^2$
minimization.  We restricted the fit to epochs ranging from 10 days
before $B$-band maximum light (the earliest points in the templates)
to 25 days after $B$-band maximum (when differences in filter
bandpasses and spectral features combine to create deviations from the
templates).  We restricted the fit to 19 days after $B$-band maximum
for the $J$ band as it showed larger deviations from the templates.
The $J$ filter in WHIRC is significantly different than the $J$ filter
used for the \citet{woodvasey08} templates (and the $K_s$ filter has
differences as well).  Details on the WHIRC filter bandpasses are
available from the WHIRC
website\footnote{http://www.noao.edu/kpno/manuals/whirc/filters.html}
and are shown compared to 2MASS and Carnegie Supernova Program
\citep[CSP; ][]{contreras10} filter bandpasses in Figure
\ref{filtfig}.

The \citet{woodvasey08} templates are defined relative to $B$-band
maximum.  We did not have optical photometry, so both the scaling in
magnitude and the epoch were free parameters in the fit.  For all
three bandpasses, we derived the same epoch of $B$-band maximum, 2011
September 9.9 $\pm$ 0.2 (MJD 55813.9, consistent with the time derived by
other groups; W. Li, private communication).

From the \citet{woodvasey08} templates we can derive the maximum in
each passband (we missed a measurement of that epoch as a result of
poor weather) as well as the magnitude in each band at the time of
$B$-band maximum.  The values at $B$-band maximum are the fiducial
points of the \citet{woodvasey08} templates.  For the maximum in each
bandpass, we find $J_{\rm max} = 10.51 \pm 0.04$ mag, $H_{\rm max} =
10.75 \pm 0.04$ mag, and $K_{s\rm max} = 10.64 \pm 0.05$ mag.  At
$B$-band maximum, we find $J_{B_{\rm max}} = 10.62 \pm 0.04$ mag,
$H_{B_{\rm max}} = 10.85 \pm 0.04$ mag, and $K_{sB_{\rm max}} = 10.68
\pm 0.05$ mag.  To evaluate the uncertainty for each value, we used
the \citet{woodvasey08} light curves that had at least three points
within three days of $B$-band maximum to calculate the error of the
mean for the template in each band.  Templates derived from data taken
by the CSP \citep{contreras10} as well as the \citet{woodvasey08} data
(Shappee \& Jha, in prep.) show essentially the same structure near
maximum brightness where we are performing the fit and result in
similar fits.

\citet{krisciunas04b} present a third-order polynomial fit for their
NIR light curves of SNe~Ia.  This curve is not a good match for our
points over the nominal range given by \citet{krisciunas04b}, most
likely as result of their dataset having few points before maximum.
If we restrict the fit to just the points near maximum, we derive
$J_{\rm max} = 10.49 \pm 0.06$ mag, $H_{\rm max} = 10.76 \pm 0.08$
mag, and $K_{s \rm max} = 10.65 \pm 0.08$ mag (uncertainties are
dominated by the rms values from \citet{krisciunas04b} fits).  All
derived magnitudes are consistent with those found using the
\citet{woodvasey08} templates.

We have not applied any $K$ corrections to our photometry.  The
redshift of M101 is $0.000804 \pm 0.000007$ \citep{devaucouleurs91},
so any $K$ correction will be minimal.  The \citet{sfd98} dust maps
imply a foreground extinction of $A_V = 0.028$ mag, and thus values in
the $J$, $H$, and $K_s$ bands less than 0.01 mag.  Based on narrow
\ion{Na}{1}~D absorption lines in the spectra of SN~2011fe,
\citet{nugent11b} derive a host-galaxy extinction of $A_V = 0.04$ mag
(\citet{patat11} report a similar result).  Again, this implies
extinctions in the $H$ and $K_s$ bands less than 0.01 mag, while the
extinction in the $J$ band is $\sim 0.01$ mag.  Given these minimal
values, we do not apply any correction to our photometry.




\section{Discussion}

To establish the absolute luminosity calibration of SNe~Ia, we need
SNe in nearby galaxies to which distances can be determined by other
methods.  There is a rich history of obtaining Cepheid-based distances
to such host galaxies, using the \emph{HST} \citep[e.g.,
][]{sandage06, gibson00, freedman01, riess09}.  A discussion of the
details associated with these techniques is beyond the scope of this
paper. Rather, we recognize that M101 provides a better platform for
the absolute magnitude calibration, because it is nearby and its
distance is better determinable, if not already determined. In
addition, the demonstration by \citet{krisciunas04a, krisciunas04b,
  woodvasey08} that the $H$-band light curves and peak brightnesses of
SNe~Ia are independent of second parameter characteristics of
individual SNe (in addition, of course to being almost unaffected by
reddening), provides a method that mitigates many of the issues that
have plagued the earlier attempts and resulted in controversy.  The
weak relationship between luminosity and decline-rate in the $J$ and
$H$ bands found by \citet{kattner12} does add some potential
complications for SNe Ia in the infra-red in general, but not for the
particular case of SN~2011fe as its $\Delta m_{15}(B)$ value of 1.2
\citep{richmond12}.  Here we use our data for SN~2011fe to discuss
the absolute magnitude anchor for the $H$-band calibration of SNe~Ia,
by comparing against currently available Cepheid distances to M101.

For the purpose of comparing our data to absolute calibrations of SNe
Ia in the infra-red, we will focus on the $H$ band, although similar
results can be found using the $J$ and $K_s$ bands.  As can be seen in
Figure \ref{filtfig}, the $H$ filter bandpass is the most similar
across the data sets used for absolute calibration (PAIRITEL and CSP)
as well as the WHIRC data.  In addition, not all derivations of
absolute calibrations include the $K_s$ band.  As an example,
\citet{woodvasey08} use $H$-band magnitudes at $B_{\rm max}$ and the
Hubble diagram (recession velocity vs apparent magnitude) to, in
effect, derive that the absolute $H$-band magnitude at $B_{\rm max}$
is:
\begin{equation}
   M( H_{B{\rm max}} ) - 5 {\rm log} (H_{0}/72)  =  -18.08 \pm 0.15 
\end{equation}
They then quote $ M( H_{B{\rm max}} ) = -18.08 \pm 0.15 $ mag, by
adopting $H_{0} = 72\ \kmsmpc$\footnote{Using only the PAIRITEL
  subsample of \citet{woodvasey08} to facilitate cross-comparisons
  with independent samples.}.
 Our measured value of  $H_{B_{\rm max}} = 10.85 \pm
0.04$ mag for SN~2011fe yields a distance modulus $(m-M)_{0} = 28.93
\pm 0.16$ mag if $H_{0}$ is indeed 
$72\ \kmsmpc$.  More precisely, SN~2011fe gives the distance modulus to M101 as:
\begin{equation}
   (m-M)  + 5 {\rm log} (H_{0}/72)  = 28.93 \pm 0.16\ \rm{mag}
\end{equation}

Note that the main uncertainty comes from the intrinsic rms of $0.15$
mag in $H$-band absolute calibration as reported by
\citet{woodvasey08}, and not from the relatively insignificant
uncertainty in the determination of the $H$ magnitude at the epoch of
$B_{max}$.  Table \ref{mods} lists the various absolute infra-red
calibrations for SNe~Ia (all essentially based on a cosmology that
assumes $H_0 = 72\ \kmsmpc$) and the distance moduli to M101 derived
from these calibrations using our apparent magnitudes for SN~2011fe.
Again, in each case the uncertainy is dominated by the absolute
calibration, not the photometry of SN~2011fe.  There is a wide range
in the absolute calibrations, yielding a span of 0.31 mag in distance
modulus depending on the specific calibration used.  The source of
this dispersion is not clear, but may be the result of different
filters, corrections to those filters, and assumptions that went into
the individual analyses.  Note also that these calibrations are not
all independent, as many use the same data sets and analysis tools.
Until the infra-red absolute magnitudes of SNe~Ia are more firmly
settled, there will be some question about their cosmological utility.

\citet{freedman01} concluded that the distance modulus to M101 from
Cepheids is $29.13 \pm 0.11$ mag, where the Cepheid distance scale
zero-point rests on an adopted LMC distance modulus ($\mu_{0}$) of
18.50 mag.  \citet{saha06} give $29.17 \pm 0.09$ mag from an
alternative analysis of the same data and an adopted $\mu_{0} (LMC)$
of 18.54 mag.  A more recent comprehensive and completely independent
study of Cepheids in M101 yields $29.04 \pm 0.19$ mag
\citep{shappee11}, where the Cepheid scale is based on the maser
distance to NGC~4258, which is tantamount to $\mu_{0} (LMC) = 18.41$
mag.  The differences among these three results thus rest entirely on
the adopted zero-point for the respective Cepheid P-L relations used
by the three sets of authors.  Comparing the conditional ($H_{0} =
72$) $H$-band distance moduli to M101 from Table \ref{mods} to the
Cepheid distances and using equation 2 (with the appropriate infra-red
calibration), one finds the infra-red distances can accommodate
$H_{0}$ values from 64 to 74 $\kmsmpc$.  Using $H$-band photometry
with NICMOS on \emph{HST}, and an assumed LMC modulus of 18.50 mag,
\citet{Macri11} obtained distance moduli of 29.53 and 29.19 mag from
Cepheids (relative to LMC Cepheids, without any metallicity dependence
modeling) in outer and inner fields of M101, respectively.  They
concluded that in addition to metallicity differences, photometry
errors from blending are a likely contributor to the observed
difference (with the inner field distance erring on the side of
appearing too close).

\begin{figure}
\epsscale{.80}
\plotone{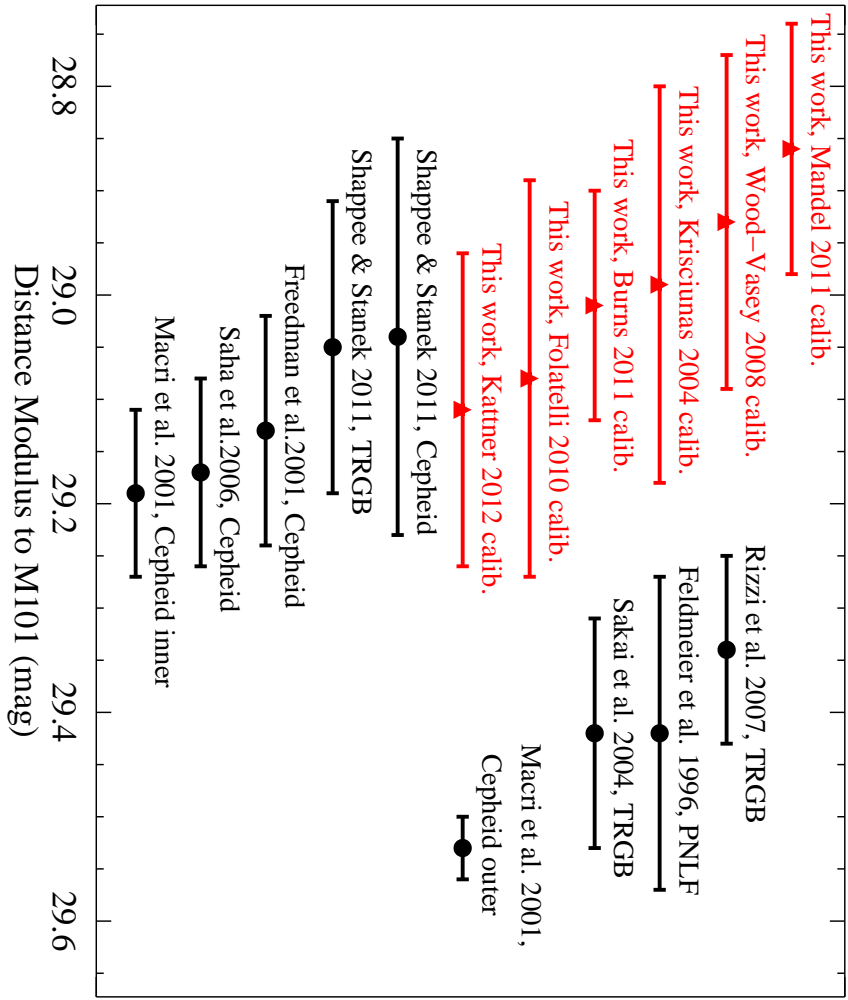}
\caption{Distance moduli to M101 derived from various techniques.
  This is not an exhaustive list of all distances to M101, but rather
  a representative sample of modern values.  The red triangles are
  distances derived from the light curve of SN~2011fe of this paper
  using the various calibrations of NIR SNe~Ia magnitudes, all of
  which assume a value of $H_{0} = 72\ \kmsmpc$. Higher values of
  $H_{0}$ slide the red points to the left, and lower values slide
  them to the right.  The black squares indicate distances derived
  from other techniques (labeled for each point).  The black points
  are independent of any assumed value for $H_{0}$, and are held
  fixed.  Error bars are 1$\sigma$.\label{distfig}}
\end{figure}

In addition, there are published distances to M101 using non-Cepheid
based methods.  Tip of the red giant branch (TRGB) results span the
gamut from $(m-M)_{0} = 29.05 \pm 0.14$ mag \citep{shappee11} to
$29.34 \pm 0.09$ mag \citep{Rizzi07} and $29.42 \pm 0.11$ mag
\citep{Sakai04}. \citet{tammann11} adopt the mean of the last 2 values
when deriving $H_{0}$ from the visible light curve of SN~2011fe.
Using the planetary nebulae luminosity function method
\citet{Feldmeier96} obtained $(m-M)_{0} =29.42 \pm 0.15$ mag.  Figure
\ref{distfig} graphically demonstrates the distance estimates for M101
discussed herein (using the $H$-band calibration for SN~2011fe).
Presented with this range of results, and given the state of the art
uncertainties in our understanding of metallicity dependence and its
inter-relation with de-reddening procedures there is no compelling
argument that can pinpoint the determined distance to M101 better than
the likely range from 29.04 to 29.42 mag.  For this range of possible
moduli and the range of conditional moduli implied by the $H$-band
magnitude of SN~2011fe, values of $H_{0}$ from 56 to 76 $\kmsmpc$
cannot be ruled out from this SN alone.

 It is sobering that M101, which is nearer than any of the SNe~Ia
 calibrating host galaxies used by \citet{freedman01} or
 \citet{Riess11}, and for which there are multiple independent
 distance determinations, has resulting distance moduli that span a
 range of $\sim$0.4 mag.  Other calibrator host-galaxies at distances
 comparable to Virgo and beyond do not offer such cross-validation to
 scrutinize the robustness of their derived distances.  The source of
 uncertainty in the distance moduli derived from the magnitudes of
 SN~2011fe is not just the measurement or the calibration of the
 SNe~Ia.  It is also necessary to resolve the Cepheid and TRGB distance
 scales and their systematics before a better than 5\% accuracy for
 $H_{0}$ can be asserted.

\section{Conclusions}

We have presented $J$, $H$, and $K_s$ light curves of SN~2011fe in
M101.  The light curves appear to be those of a normal SN Ia.  Our
apparent magnitude in the $H$ band at the epoch of $B$-band maximum is
$10.85 \pm 0.04$, implying distance moduli to M101 based on various
infra-red absolute calibrations that span a range from 28.86 to 29.17
mag.  This dispersion is comparable to that for traditional distance
measures to M101 (29.04 to 29.42 mag).  This is, however, only one
object in a class that still exhibits a small, but significant,
intrinsic spread in peak magnitudes.  From the dispersion in absolute
calibrations of SNe~Ia in the infra-red, it is clear that they are not
yet fully understood.

\acknowledgments

We would like to thank the referee, Mark Phillips, for extremely
useful comments and suggestions.  We would also like to thank the WIYN
Observatory for their support of this program.  T.M. acknowledges many
useful conversations with Chris Burns on the nature of SN light curves
in the infra-red.  T.M. dedicates this paper to the memory of his
friend and colleague, Dr. Weidong Li.

{\it Facilities:} \facility{WIYN}

\clearpage

\clearpage

\begin{deluxetable}{cccc}
\tablecaption{Apparent magnitudes \label{irphot}}
\tablewidth{0pt}
\tablehead{
\colhead{MJD} & \colhead{$J$}  & \colhead{$H$} &
 \colhead{$K_s$} 
}
\startdata
55800.121 & 12.860 (010) & 12.822 (009) & 12.855 (031) \\ 
55802.168 & 11.901 (010) & 11.898 (008) & 11.967 (029) \\ 
55803.121 & 11.566 (011) & 11.587 (008) & 11.609 (012) \\ 
55804.195 &    ...       & 11.317 (010) &    ...       \\ 
55805.199 &    ...       & 11.124 (008) &    ...       \\ 
55807.148 & 10.685 (013) & 10.849 (008) & 10.809 (014) \\ 
55808.102 & 10.583 (010) & 10.768 (008) & 10.720 (011) \\ 
55813.113 & 10.503 (009) & 10.804 (008) & 10.641 (013) \\ 
55819.133 & 10.968 (009) & 11.028 (008) & 10.890 (012) \\ 
55820.090 & 11.072 (009) & 11.036 (008) & 10.912 (011) \\ 
55821.090 & 11.224 (012) & 11.035 (009) & 10.955 (012) \\ 
55822.090 & 11.400 (010) & 11.059 (008) & 10.994 (013) \\ 
55824.090 & 11.860 (009) & 11.096 (008) & 11.061 (011) \\ 
55825.094 & 12.063 (009) & 11.100 (008) & 11.076 (012) \\ 
55827.121 & 12.326 (010) & 11.084 (008) & 11.075 (012) \\ 
55828.086 & 12.382 (011) & 11.045 (009) & 11.061 (013) \\ 
55830.109 & 12.394 (009) & 10.966 (008) & 11.014 (012) \\ 
55833.090 & 12.287 (011) & 10.827 (008) & 10.909 (012) \\ 
55834.098 &    ...       & 10.779 (015) &    ...       \\ 
55835.090 & 12.200 (010) & 10.771 (008) & 10.861 (014) \\ 
55837.086 & 12.111 (010) & 10.726 (008) & 10.821 (012) \\ 
55838.078 & 12.051 (010) & 10.715 (008) & 10.808 (013) \\ 
55840.070 & 11.927 (009) & 10.704 (008) & 10.818 (011) \\ 
55842.074 & 11.788 (011) & 10.734 (008) & 10.845 (011) \\ 
55844.074 & 11.713 (010) & 10.797 (008) & 10.939 (012) \\ 
55845.105 &    ...       & 10.845 (009) &    ...       \\ 
55846.094 & 11.718 (010) & 10.891 (009) & 11.040 (015) \\ 
55848.074 & 11.865 (022) & 11.030 (008) & 11.215 (013) \\ 
55849.070 & 11.955 (011) & 11.101 (008) & 11.289 (012) \\ 
55850.066 & 12.052 (009) & 11.178 (010) & 11.405 (026) \\ 
55851.066 & 12.135 (016) & 11.245 (008) & 11.467 (013) \\ 
55852.070 & 12.257 (013) & 11.304 (008) & 11.517 (013) \\ 
55853.062 & 12.338 (011) & 11.356 (010) & 11.569 (015) \\ 
55854.070 & 12.425 (012) & 11.415 (010) & 11.632 (013) \\ 
55860.070 & 12.920 (011) & 11.708 (008) & 11.927 (012) \\

\enddata

\end{deluxetable}
\begin{deluxetable}{lccc}
\tablecaption{Derived Distance Moduli of M101\label{mods}}
\tablewidth{0pt}
\tablehead{
\colhead{Calibration Source} & \colhead{Filter}  & \colhead{Absolute Magnitude} &
 \colhead{Distance modulus} \\
\colhead{} & \colhead {} & \colhead{} & \colhead{to M101 (mag)\tablenotemark{a}}
}
\startdata
\citet{mandel09}\tablenotemark{b} & $J$   & -18.25 $\pm$ 0.17 & 28.87 $\pm$ 0.17 \\ 
                 & $H$   & -18.01 $\pm$ 0.11 & 28.86 $\pm$ 0.12 \\ 
                 & $K_s$ & -18.25 $\pm$ 0.19 & 28.93 $\pm$ 0.20 \\ 
\tableline
\citet{woodvasey08}\tablenotemark{b,c} & $J$ &  -18.29 $\pm$  0.09 &  28.91 $\pm$  0.10 \\ 
      &  $H$ & -18.08 $\pm$  0.15 &  28.93 $\pm$  0.16 \\ 
       & $K_s$ & -18.32 $\pm$  0.26 &  29.00  $\pm$  0.26 \\ 
\tableline
\citet{krisciunas04a}\tablenotemark{d} & $J$ &  -18.57 $\pm$  0.14 & 29.08 $\pm$  0.14 \\ 
     &  $H$ &  -18.24 $\pm$  0.18 & 28.99 $\pm$  0.19 \\ 
      &  $K_s$ &  -18.42 $\pm$  0.12  & 29.06 $\pm$  0.13 \\ 
\tableline
\citet{folatelli10}\tablenotemark{b} & $J$  & -18.42 $\pm$  0.18 &  29.04  $\pm$ 0.18  \\ 
     &   $H$  & -18.23 $\pm$ 0.19  & 29.08 $\pm$ 0.19 \\ 
      &  $K_s$ & -18.30 $\pm$ 0.27 &  28.98 $\pm$ 0.27 \\ 
\tableline
\citet{folatelli10}\tablenotemark{d} & $J$  & -18.43 $\pm$  0.18 & 28.94 $\pm$  0.18 \\ 
     &   $H$  & -18.42 $\pm$  0.19 & 29.17 $\pm$  0.19 \\ 
      &  $K_s$ & -18.47 $\pm$  0.27 & 29.11 $\pm$  0.27 \\
\tableline 
\citet{burns11}\tablenotemark{d}  & $J$ & -18.44 $\pm$  0.12 & 28.94 $\pm$ 0.13 \\ 
      &  $H$ & -18.26 $\pm$ 0.10 & 29.01 $\pm$ 0.11 \\ 
\tableline
\citet{kattner12}\tablenotemark{d,e} &  $J$  & -18.57 $\pm$  0.14  & 29.08 $\pm$  0.15 \\ 
     &   $H$  & -18.42 $\pm$  0.14 & 29.17 $\pm$  0.15 \\ 

\enddata
\tablenotetext{a}{Distance modulus calculated by combining absoluted
  magnitude with the apparent magnitudes for SN~2001fe derived from
  our light curves (see text for details).  This is conditional on
  $H_{0}$ = 72 $\kmsmpc$.}
\tablenotetext{b}{Fiducial time is $B$-band maximum brightness.}
\tablenotetext{c}{Using PAIRITEL subsample only.}
\tablenotetext{d}{Fiducial time is maximum brightness in the given
  filter ($J$, $H$, or $K_s$).}
\tablenotetext{e}{Using subsample 2 of \citet{kattner12}.}

\end{deluxetable}

\clearpage

\end{document}